\documentclass[showpacs,preprintnumbers,amsmath,amssymb]{revtex4}

\usepackage[all]{xy}
\usepackage{graphicx}
\usepackage{dcolumn}
\usepackage{bm}
\usepackage{rotating}

\newtheorem{theorem}{Theorem}

\newtheorem{definition}{Definition}
\newtheorem{example}{Example}

\newtheorem{lemma}{Lemma}

\newtheorem{proposition}{Proposition}
\newtheorem{remark}{Remark}

\newcommand{\beq}{\begin{equation}}
\newcommand{\eeq}{\end{equation}}
\newcommand{\beqa}{\begin{eqnarray}}
\newcommand{\eeqa}{\end{eqnarray}}
\newcommand{\noi}{\noindent}

\newcommand{\g}{{\mathfrak g}}
\newcommand{\h}{{\mathfrak h}}

\newcommand{\n}{{\mathfrak n}}

\def\>{\rangle}
\def\<{\langle}

\begin{document}

\title{Hopf algebras for ternary algebras }

\author{M. Goze}
\email{m.goze@uha.fr}
\affiliation{%
Laboratoire MIA, Universit\'e de Haute Alsace, \\
Facult\'e des Sciences et Techniques, 4 rue des Fr\`eres
Lumi\`ere, 68093 Mulhouse Cedex, France
}%
\author{M. Rausch de Traubenberg}
\email{Michel.Rausch@IReS.in2p3.fr}
\affiliation{
IPHC-DRS, UdS, CNRS, IN2P3; 23  rue du Loess, 67037 Strasbourg, France
}

\date{\today}

\begin{abstract}
We construct an universal enveloping algebra associated to the 
ternary extension of Lie (super)algebras called Lie algebra of order
three. A  Poincar\'e-Birkhoff-Witt theorem is proven is this context.
It this then shown that this universal enveloping algebra can be endowed
with a structure of Hopf algebra.
 The study of the dual of the universal
enveloping algebra enables to define the parameters of the transformation
of a Lie algebra of order three. It turns out that these 
variables  are the variables which
generate the three-exterior algebra.  
\end{abstract}

\pacs{02.10.Xm, 02.20.Sv, 11.30.Ly}
\keywords{Ternary algebras and groups, Hopf algebra, three-exterior algebra }
\maketitle

\section{Introduction}
The concept of algebras, that is a vector space  ${\cal A}$ equipped
with a binary product $m_2 \ : \ {\cal A} \otimes {\cal A} \longrightarrow
{\cal A}$ is  central in physics. Among them Lie  algebras and Lie 
superalgebras are the  cornerstone in the construction of models in
particles physics. 
The former lead to  a description of space-time and internal symmetries
although the latter give rise to supersymmetric extensions
of space-time symmetries.
The intensive use of  Lie (super)algebras
 is probably due to the central theorems
of Coleman \& Mandula \cite{cm} and Haag, Lopuszanski \&
Sohnius \cite{hls} which allow to construct theories based on some
specific Lie (super)algebras  not
contradicting the principles of Quantum Theory and Relativity.

On may naturally wonder  whether or not some different algebraic
structures should play a role in physics. And in particular
one may ask if ternary algebras, that is  vector spaces
${\cal A}$ equipped with a ternary multiplication: $m_3 \ : \
{\cal A} \otimes {\cal A} \otimes {\cal A} \longrightarrow {\cal A}$
should be relevant in the description of some symmetries. 
Ternary algebras have been considered in physics only occasionally
(see for instance \cite{bg1,bg2,k1,k2,k3,r} and references therein).
For some mathematical references one can see 
\cite{f,g,mv,gr}. Recently they was some revival of interest in 
ternary algebras when it has been realised that a
 ternary algebra defined by a fully
antisymmetric product appears in the description
of  multiple M2-branes \cite{bl}.

In \cite{flie1,flie2,flie3} an   
$F-$ary algebra which can be seen as a possible
generalisation of Lie (super)algebras has been considered and named
Lie algebra of order $F$. 
A Lie algebra of order $F$  admits a $\mathbb Z_F-$grading
($F=3$ in this paper), the zero-graded part
being a Lie algebra. An $F-$fold symmetric product (playing the role of
the anticommutator in the case $F=2$)  expresses the zero graded  part
in terms of the non-zero graded part. This means that when $F=3$ we have two
products: one binary and the second ternary.
The first algebras  constructed along these lines
 lead to some 
(infinite-dimensional) extensions
of the Poincar\'e algebra in $(1+2)-$dimensions  and turn out to induce
a symmetry which connects relativistics anyons \cite{anyons}. 
A major progress in the comprehension of those mathematical structures
was undertaken when it was realised that finite dimensional Lie
algebras of order $F$ could be defined. 
 Subsequently, a  specific (finite-dimensional) Lie algebra or order three,
leading to a non-trivial extension of the Poincar\'e algebra,
  has been studied 
together with its implementation in Quantum Field Theory  
\cite{cubic1,cubic3,pform,noether}. 
Then, a general study of the possible non-trivial extensions of the
Poincar\'e algebra in $(1+3)-$dimensions 
has been undertaken in \cite{flie3,grt1} together with a  study
of possible  kinematical algebras of order three \cite{cr}.

However, all these mathematical structures have been considered at
the level of  algebras {\it i.e.} at the level of infinitesimal
transformations and no groups associated to Lie algebras of order three
were considered. At a first glance these  two structures seem to
be incompatible since for a Lie algebra of order three for some elements
only the product of three elements is defined although for a group
the product of two elements is always defined.  This is rather different
to the Lie (super)algebras cases where it is known that Lie (super)groups
can be defined. However, on the formal ground it is knows that
Lie (super)groups are related to Hopf algebras.
Indeed, if $G$ is a simply connected
Lie group, there is a duality between ${\cal U}(\g)$ (the universal
enveloping algebra of $\g$, the Lie algebra of $G$) and ${\cal F}(G)$
(the vector space of complex valued functions on $G$). This means that 
${\cal F}(G)$ is isomorphic to a subspace of ${\cal U}^*(\g)$ (the dual
of ${\cal U}(\g)$)
\cite{cp,majid}.

The purpose of this paper is to give a first step toward a construction
of groups associated to Lie algebras of order three. We will show that
 it is possible  to define
an universal enveloping algebra associated to a Lie algebra of order three,
and then to endow it with a structure of Hopf algebra.

The content of the paper is the following. In section two the definition
of a Lie algebra of order $F$, together with some explicit examples are
given. It is  shown in section three that one can define a universal
enveloping algebra associated to a Lie algebra of order three, and
a Poincar\'e-Birkhoff-Witt theorem is established.
It turns out that the  PBW basis 
is strongly related to the three-exterior algebra \cite{roby}. 
In section four the universal enveloping algebra is endowed with
a Hopf algebra structure. 
 In section five
the study of the Hopf dual   enables us to define 
the parameters of the transformation which 
turn out to be related to  the three-exterior algebra.
A conclusion 
 is given in Section VI.
Some of the results of this paper was partially given in \cite{valla}.

It has to be noticed that some attempt to consider a group
associated to a Lie algebra of order three
$\g$  has been undertaken in \cite{ayu},
where they have endowed ${\cal U}(\g)$ with a Hopf algebra structure and
 they identified the dual of ${\cal U}(\g)$ with the group
associated with $\g$. 
Their construction of Hopf algebra structure on ${\cal U}(\g)$
is slightly different from ours since they are considering untwisted Hopf
algebra although we are considering twisted Hopf algebras (see Section 4.).
They also obtained a result similar to Proposition \ref{dual-prod}
which enable them to define a group associated to a Lie algebra of
order three considering specific polynomials.
However, even if
their construction is similar to our approach,
 there is no description of a basis of ${\cal U}(\g)$ and the construction of the dual is incomplete.

\section{Lie algebras of order $F$}

In this section we recall the definition and some basic properties of Lie
algebras of order $F$ introduced in \cite{flie1, flie2}. Some examples
useful for the sequel are also given.

\subsection{Definition and examples of elementary Lie algebras of order three}

We consider ${\mathfrak{g}}$ a complex vector space and $\varepsilon$ an
automorphism of ${\mathfrak{g}}$ satisfying $\varepsilon^F=1$. The vector
space ${\mathfrak{g}}$ is decomposed in ${\mathfrak{g}}={\mathfrak{g}}%
_0\oplus {\mathfrak{g}}_1 \cdots \oplus {\mathfrak{g}}_{F-1}$ where ${%
\mathfrak{g}}_i$ is the eigenspace corresponding to the eigenvalue $q^i$ of $%
\varepsilon$ where $q=e^{2i\pi/F}$.

\begin{definition}
Let $F\in \mathbb{N}^{\ast }$. A ${\mathbb{Z}}_{F}$-graded ${\mathbb{C}}-$%
vector space ${\mathfrak{g}}={\mathfrak{g}}_{0}\oplus {\mathfrak{g}}%
_{1}\oplus {\mathfrak{g}}_{2}\dots \oplus {\mathfrak{g}}_{F-1}$ is called a
complex Lie algebra of order $F$ if

\begin{enumerate}
\item $\mathfrak{g}_0$ is a complex Lie algebra.

\item For all $i= 1, \dots, F-1 $, $\mathfrak{g}_i$ is a representation of $%
\mathfrak{g}_0$. If $X \in {\mathfrak{g}}_0, \ Y \in {\mathfrak{g}}_i$ then $%
[X,Y]$ denotes the action of $X$ on $Y$ for any $i=1,\cdots,F-1$.

\item For all $i=1,\dots,F-1$, there exists an $F-$linear, $\mathfrak{g}_0-$%
equivariant map 
\begin{equation*}
\{ \cdots \} : \mathcal{S}^F\left(\mathfrak{g}_i\right) \rightarrow 
\mathfrak{g}_0,
\end{equation*}
where $\mathcal{S}^F(\mathfrak{g}_i)$ denotes the $F-$fold symmetric product
of $\mathfrak{g}_i$, satisfying the following (fundamental) identity 
\begin{eqnarray}  \label{eq:J}
&&\sum\limits_{j=1}^{F+1} \left[ \left\{ Y_1,\dots, Y_{j-1},
Y_{j+1},\dots,Y_{F+1}\right\},Y_i \right] =0,  
\end{eqnarray}
\noindent for all $Y_j \in \mathfrak{g}_i$, $j=1,..,F+1$.
\end{enumerate}
\end{definition}

\begin{remark}
If $F=1$, by definition $\mathfrak{g}=\mathfrak{g}_0$ and a Lie algebra of
order one is a Lie algebra. If $F=2$, then ${\mathfrak{g}}$ is a Lie
superalgebra. Therefore, Lie algebras of order $F$ appear as some kind of
generalisations of Lie algebras and superalgebras. 
\end{remark}

\noindent
It is important to notice
that the bracket 
 $\{ \cdots \}$ is a priori  not defined for  elements in different
gradings. This leads to the following definition.

\begin{proposition}
Let ${\mathfrak{g}}={\mathfrak{g}}_0\oplus{\mathfrak{g}}_1\oplus\dots\oplus{%
\mathfrak{g}}_{F-1}$ be a Lie algebra of order $F$, with $F>1$. For any $%
i=1,\ldots,F-1$, the ${\mathbb{Z}}_F-$graded vector spaces ${\mathfrak{g}}%
_0\oplus{\mathfrak{g}}_i$ is a Lie algebra of order $F$. We call these type
of algebras \textit{elementary Lie algebras of order $F$}.
\end{proposition}

In \cite{flie2} an inductive process for the construction of Lie algebras of
order $F$ starting from a Lie algebra of order $F_1$ with $1\le F_1<F$ is
given. In this paper, we restrict ourselves to elementary Lie algebras of
order three, ${\mathfrak{g}}={\mathfrak{g}}_0\oplus{\mathfrak{g}}_1$.
Non-trivial examples of Lie algebras of order $F$ (finite and
infinite-dimensional) are given in \cite{flie1, flie2}. We now give some
examples of finite-dimensional Lie algebras of order three, which will be
relevant in the sequel.

\begin{example}
\label{so23} 
Let $\g$ be any finite-dimensional Lie algebra provided with
a symmetric invariant bilinear form $(\,,\,): \g \otimes \g \to \mathbb C$
and define $\g_0=\g, \g_1=\g$ with
adjoint representation and $\{\,,\,,\,\}$ by 
$$ \{Y_1, Y_2, Y_3\} = (Y_1, Y_2) Y_3 +(Y_2, Y_3) Y_1 +(Y_1, Y_3) Y_2 ,$$
this is  a Lie algebra of order three. 
In particular, if 
${\mathfrak{g}_0}$
 is a semi-simple Lie algebra and ${%
\mathfrak{g}}_1=\text{ad}(\g_0)$ with
  $\{J_a, a =1,\cdots, \text{%
dim}\ {\mathfrak{g}}_0\}$  a basis of ${\mathfrak{g}}_0$ and $\{A_a, a
=1,\cdots, \text{dim}\ {\mathfrak{g}}_0\}$  the corresponding basis of ${%
\mathfrak{g}}_1$ we have 

\begin{equation}
[J_a,J_b]=f_{ab}{}^c J_c, \ \ [J_a,A_b]=f_{ab}{}^c A_c, \ \ \{ A_a,A_b,A_c
\}=g_{ab}J_c+g_{ac}J_b+g_{bc}J_a,  \notag
\end{equation}

where   $g_{ab}=Tr(A_aA_b)$ is  the Killing form and $%
f_{ab}{}^c $  the structure constants of ${\mathfrak{g}}_0$.

\end{example}

\begin{example}
\label{FP} Let ${\mathfrak{g}}_0 = \left< L_{\mu \nu }=-L_{\nu \mu}, P_\mu,
\ \mu,\nu=0,\cdots,D-1\right>$ be the Poincar\'e algebra in $D-$dimensions
and ${\mathfrak{g}}_1=\left<V_\mu, \ \mu=0,\cdots, D-1 \right>$ be the $D-$%
dimensional vector representation of ${\mathfrak{g}}_0$. The brackets 
\begin{eqnarray}  \label{fp-bracket}
\left[L_{\mu \nu }, L_{\rho \sigma}\right]&=& \eta_{\nu \sigma } L_{\rho \mu
}-\eta_{\mu \sigma} L_{\rho \nu} + \eta_{\nu \rho}L_{\mu \sigma} -\eta_{\mu
\rho} L_{\nu \sigma},  \notag \\
\left[L_{\mu \nu }, P_\rho \right]&=& \eta_{\nu \rho } P_\mu -\eta_{\mu \rho
} P_\nu, \ \left[L_{\mu \nu }, V_\rho \right]= \eta_{\nu \rho } V_\mu
-\eta_{\mu \rho } V_\nu, \ \left[P_{\mu}, V_\nu \right]= 0,  \notag \\
\{ V_\mu, V_\nu, V_\rho \}&=& \eta_{\mu \nu } P_\rho + \eta_{\mu \rho }
P_\nu + \eta_{\rho \nu } P_\mu,  \notag
\end{eqnarray}
\noindent with the metric $\eta_{\mu \nu}=\mathrm{{diag}(1,-1,\cdots,-1)}$
endow ${\mathfrak{g}}={\mathfrak{g}}_0 \oplus {\mathfrak{g}}_1$ with an
elementary Lie algebra of order three structure which is denoted ${\mathfrak{%
iso}}_3(1,D-1)$.
\end{example}

It has been shown that Example \ref{so23} (with ${\mathfrak{g}}_0=\mathfrak{%
so}(2,3)$) and Example \ref{FP} (when $D=4$) are related through an
In\"on\"u-Wigner contraction \cite{flie2} and a deformation in a
Gerstenhaber sense \cite{flie3}.

\begin{example}
Let $A$ be an associative algebra. We denote by $L(A)$ the Lie algebra
defined from $A$ by the bracket $[a_1,a_2]=a_1a_2-a_2a_1$. We consider the
graded vector space $L_3(A)=L(A) \oplus A.$ We define the following products 
\begin{equation*}
\left\{ 
\begin{array}{l}
[a_1,a_2]=a_1a_2-a_2a_1 \ \in L(A), \ \ \forall a_1,a_2 \in L(A) \\ 
\left[a,b\right]=ab-ba \ \in A, \ \ \forall a \in L(A), \ \forall b \in A \\ 
\{b_1,b_2,b_3\}=\Sigma _{\sigma \in S_3}b_{\sigma(1)}b_{\sigma (2)}b_{\sigma
(3)} \ \in L(A), \ \ \forall b_1,b_2,b_3 \in A.%
\end{array}
\right.
\end{equation*}
It is easy to prove that these products provide $L_3(A)$ with a structure of
elementary Lie algebra of order three. 
Indeed, one can check that the fundamental identity \eqref{eq:J} is just
a consequence of the associativity of the product in $A$.
We call this algebra the elementary Lie
algebra of order three associated to the associative algebra $A$.
\end{example}

Given $A$ and $B$ to associative algebras,
notice that  the application $A \to L_3(A)$ is functorial.
Indeed, a morphism of associative algebras 
$A \to B$ induces a morphism from $L_3(A) \to L_3(B)$.
Then, given a Lie algebra of order 3 $\g$ 
and  an associative algebra $A$, there exists 
a morphism of Lie algebras of order three $\g \to L_3(A)$ if and only if
there exists a linear map $\g \to A$ satisfying the conditions in 
Definition 2. 
This remark justifies the definition of representation below.

\subsection{Morphisms and representations of an elementary Lie algebras of 
order $F$} 

Let ${\mathfrak{g}}$ and $\mathfrak{h}$ be two Lie algebras of order $F$. A
linear map $f: {\mathfrak{g}}\longrightarrow \mathfrak{h}$ is a morphism of
algebras of order $F$ if $f$ is graded, that is $f=f_0+f_1+ \cdots +f_{F-1}$
with $f_a({\mathfrak{g}}_a)\subset \mathfrak{h}_a$ for all $a \in \{0,1,
\cdots, F-1\}$ and 
\begin{equation*}
\left\{ 
\begin{array}{l}
f_a[X,X_a]=[f_0(X),f_a(X_a)], \ \ \forall \ X \in {\mathfrak{g}}_0, \ X_a
\in {\mathfrak{g}}_a \\ 
f_0\{Y_1,\cdots,Y_F\}=\{f_{i}(Y_1),\cdots,f_{i}(Y_F)\}, \ \ \forall \
Y_1,\cdots,Y_F \in {\mathfrak{g}}_{i}, \ i=1,\cdots, F-1.%
\end{array}
\right.
\end{equation*}
Let $V$ be a $\mathbb{Z}_F-$graded vector space.

\begin{definition}
\label{representation} A representation of an elementary Lie algebra of
order $F$ is a linear map $\rho : ~ {\mathfrak{g}}={\mathfrak{g}}_0 \oplus {%
\mathfrak{g}}_1 \to \mathrm{End}(V)$, such that (for all $X_i \in {\mathfrak{%
g}}_0, Y_j \in {\mathfrak{g}}_1$)

\begin{eqnarray}  \label{eq:rep}
\begin{array}{ll}
& \rho\left(\left[X_1,X_2\right]\right)= \rho(X_1) \rho(X_2)-
\rho(X_2)\rho(X_1) \cr & \rho\left(\left[X_1,Y_2\right]\right)= \rho(X_1)
\rho(Y_2)- \rho(Y_2)\rho(X_1) \cr & \rho\left(
\left\{Y_1.\cdots,Y_F\right\}\right)= \sum \limits_{\sigma \in S_F}
\rho\left(Y_{\sigma(1)}\right) \cdots \rho\left(Y_{\sigma(F)}\right) \cr%
\end{array}%
\end{eqnarray}

\noindent ($S_F$ being the group of permutations of $F$ elements).
\end{definition}

\noindent If $V= V_0 \oplus \cdots \oplus V_{F-1}$ then for all $a
\in\{0,\cdots, F-1\}$, $V_a$ is a ${\mathfrak{g}}_0-$module and we have $%
\rho({\mathfrak{g}}_1) (V_a) \subseteq V_{a+1}$.


\medskip

\begin{example}
\label{mat} Let $\g(m_1,m_2,m_3)$ and $\g_{\text{el}}(m_1,m_2,m_3)$ 
be the set of $(m_1 + m_2 +m_3) \times (m_1 + m_2 + m_3)$
matrices of the form

\begin{eqnarray}  \label{gl}
\begin{array}{ll}
\g_{\mathrm{el}}(m_1,m_2,m_3) = \left\{ 
\begin{pmatrix}
a_0 & b_1 & 0 \\ 
0 & a_1 & b_2 \\ 
b_0 & 0 & a_2%
\end{pmatrix}
\right\}, & \g(m_1,m_2,m_3)=\left\{ 
\begin{pmatrix}
a_0 & b_1 & c_2 \\ 
c_0 & a_1 & b_2 \\ 
b_0 & c_1 & a_2%
\end{pmatrix}%
\right\},%
\end{array}%
\end{eqnarray}
\noindent with $a_0 \in \mathfrak{gl}(m_1), a_1 \in \mathfrak{gl}(m_2),a_3
\in \mathfrak{gl}(m_3),$ $b_1 \in \mathcal{M}_{m_1,m_2}(\mathbb{C}), b_2\in 
\mathcal{M}_{m_2,m_3}(\mathbb{C}), b_0 \in \mathcal{M}_{m_3,m_1}(\mathbb{C}) 
$, and $c_0 \in \mathcal{M}_{m_2,m_1}(\mathbb{C}), c_1\in \mathcal{M}%
_{m_3,m_2}(\mathbb{C}), c_2 \in \mathcal{M}_{m_1,m_3}(\mathbb{C})$.
Define $\g_0=\mathfrak{gl}(m_1)\oplus \mathfrak{gl}(m_2) \oplus
 \mathfrak{gl}(m_3)$, $\g_1= \mathcal{M}_{m_1,m_2}(\mathbb{C}) \oplus 
\mathcal{M}_{m_2,m_3}(\mathbb{C}) \oplus \mathcal{M}_{m_3,m_1}(\mathbb{C})$
and $\g_2= \mathcal{M}_{m_1,m_3}(\mathbb{C}) \oplus
\mathcal{M}_{m_3,m_2}(\mathbb{C}) \oplus
\mathcal{M}_{m_2,m_1}(\mathbb{C}) $.
It is obvious that $\g_1$ and $\g_2$ are representations of $\g_0$.
Furthermore since  $\g_i \g_j \subseteq \g_{i+j}$
we have $\g_1 \g_1 \g_1 \subseteq \g_0$ and 
$\g_2 \g_2 \g_2 \subseteq \g_0$ and consequently 
$\g(m_1,m_2,m_3)$ (resp. $\g_{\text{el}}(m_1,m_2,m_3)$)
defines a Lie algebra of order three 
(resp. an elementary Lie algebra of order three). 
\end{example}

\section{Universal enveloping algebra of Lie algebras of order three}

\label{hopf}

In this section we construct the universal enveloping algebra associated to 
an elementary Lie algebra of order three. We also show that some 
 Poincar\'e-Birkhoff-Witt theorem  holds true in this context.

\subsection{Enveloping algebra associated to an elementary Lie algebra of
order three}

We consider ${\mathfrak{g}}={\mathfrak{g}}_0 \oplus {\mathfrak{g}}_1$ an
elementary Lie algebra of order three and denote generically by $X$ (resp. $Y$%
) the elements of ${\mathfrak{g}}_0$ (resp. ${\mathfrak{g}}_1$).

\begin{definition}
The universal enveloping algebra $\mathcal{U}({\mathfrak{g}})$ is the
quotient of the tensor algebra $T({\mathfrak{g}})$ by the two sided-ideal
generated by

\begin{eqnarray}
&&X_1\otimes X_2 - X_2 \otimes X_1 -[X_1,X_2],  \notag \\
&&X_1\otimes Y_2 - Y_2 \otimes X_1 -[X_1,Y_2],  \notag \\
&&Y_1\otimes Y_2 \otimes Y_3 + Y_3\otimes Y_1 \otimes Y_2 + Y_2\otimes Y_3
\otimes Y_1 + Y_1\otimes Y_3 \otimes Y_2 + Y_3\otimes Y_2 \otimes Y_1 +
Y_2\otimes Y_1 \otimes Y_3 - \left\{Y_1,Y_2,Y_3 \right\}.  \notag
\end{eqnarray}
\end{definition}

\noi
As before, 
 the fundamental identity
\eqref{eq:J} is a direct consequence of the associativity
of the tensorial product in  $T(\g)$.
We now state the universal property of $\mathcal{U}({\mathfrak{g}})$.

\begin{theorem}
\label{univ}
Let ${\mathfrak{g}}$ be an elementary Lie algebra of order three. Given any
associative algebra $A$ and any morphism $f$ of Lie algebras of order three
from ${\mathfrak{g}}$ to $L_3(A)$, there exists an unique morphism of
algebras $\varphi : {\mathcal{U}}({\mathfrak{g}})\longrightarrow A$ such
that $\varphi \circ i_{{\mathfrak{g}}}=f$ where $i_{{\mathfrak{g}}}$ is the
composition of the canonical injection of ${\mathfrak{g}}$ into $T({%
\mathfrak{g}})$ and the canonical surjection of $T({\mathfrak{g}})$ into $%
\mathcal{U}({\mathfrak{g}})$.
\end{theorem}

{\bf Proof } By definition of the tensor algebra, $f$ extends to a
morphism of algebras $\tilde f$ from $T({\mathfrak{g}})$ to $A$ defined by $%
\tilde f(X_1 \otimes X_2)=f_0(X_1) f_0(X_2), \tilde f(X_1 \otimes
Y)=f_0(X_1)f_1(Y), {\ etc.},$ for $X_1, X_2 \in {\mathfrak{g}}_0$ and $Y \in 
{\mathfrak{g}}_1$. This map satisfies 
\begin{equation*}
\tilde f(X \otimes X^{\prime }-X^{\prime }\otimes X - [X,X^{\prime
}])=[f_0(X),f_0(X^{\prime })]-f_0[X,X^{\prime }]=0, \ \forall X,X^{\prime
}\in {\mathfrak{g}}_0
\end{equation*}
because $f_0$ is morphism of Lie algebras. Likewise we have 
\begin{equation*}
\tilde f(X \otimes Y-Y \otimes X - [X,Y])=[f_0(X),f_1(Y)]-f_1[X,Y]=0,\
\forall X \in {\mathfrak{g}}_0,Y \in {\mathfrak{g}}_1
\end{equation*}
and 
\begin{equation*}
\tilde f(\Sigma Y_1\otimes Y_2 \otimes
Y_3-\{Y_1,Y_2,Y_3\})=\{f_1(Y_1),f_1(Y_2),f_1(Y_3)\}-f_0\{Y_1,Y_2,Y_3\}=0
\end{equation*}
for any $Y_1,Y_2,Y_3 \in {\mathfrak{g}}_1$ where $\Sigma Y_1\otimes Y_2
\otimes Y_3$ is the symmetric product. Then $\tilde f$ is trivial on the
ideal generated by the relation defining $\mathcal{U}({\mathfrak{g}})$. This
proves the existence of $\varphi $. The uniqueness is due to the fact that ${%
\mathfrak{g}}$ generates the algebra $T({\mathfrak{g}})$. Q.E.D.

\subsection{The Poincar\'e-Birkhoff-Witt theorem}
Having defined the universal enveloping algebra, we now establish the
Poincar\'e-Birkhoff-Witt theorem. However, it is necessary to recall
firstly some results concerning the $n-$exterior algebra, that becomes
central in this context.\\
\subsubsection{ The $n$-exterior Roby algebra}
\label{ext} 
Let $n\ge 2$ be an integer.
The $n-$exterior algebra has been introduced by Roby \cite{roby}
(see also \cite{r,cliff}). In this section we recall some of the main results
of \cite{roby} useful for this paper. Let $V$ be a $d-$dimensional vector
space over the field $\mathbb{C}$ (or $\mathbb{R}$) and consider $T(V) = 
\mathbb{C }\oplus V \oplus (V\otimes V) \oplus V^{\otimes^3} \oplus \cdots$
the tensor algebra over $V$. Consider $v_1, v_2,\cdots, v_p$, $p$ different
independent elements of $V$ and $n_1, \cdots, n_p \in \mathbb{N}^*$ 
such that $%
n_1 + \cdots n_p =n$. Define the tensor $\sigma = \mathcal{S}%
(v_1^{\otimes^{n_1}}\otimes \cdots \otimes v_p^{\otimes^{n_p}})$, with $%
\mathcal{S}$ being the symmetrised tensorial product of the tensor $%
v_1^{\otimes^{n_1}}\otimes \cdots \otimes v_p^{\otimes^{n_p}}$. This
symmetrised tensor contains $\frac{n !}{n_1 ! \cdots n_p !} $ monomials. For
instance $\mathcal{S}(v_1^{\otimes^2}\otimes v_2)= v_1^{\otimes^2} \otimes
v_2 + v_1 \otimes v_2 \otimes v_1 + v_2 \otimes v_1^{\otimes^2}$. We
consider $\mathcal{I}(V,n)$ the two-sided ideal generated by elements of the
form $\sigma$.

\begin{definition}
\label{roby-def} Let $V$ be a $d-$dimensional vector space and let $\mathcal{%
I}(V,n)$ be defined as above. The $n-$exterior algebra is the $\mathbb{Z}_n-$%
graded vector space $\Lambda(V,n)=T(V)/\mathcal{I}(V,n).$
\end{definition}

\begin{remark}
\label{roby-ex} The composition of the natural map $V \to T(V)$ with the
canonical projection $T(V)\to \Lambda(V,n)$ gives $V \subset \Lambda(V,n)$
and we identify $V$ with its image under this map. Thus, if we denote $\{
e_i, 1 \le i \le d\}$ a basis of $V$ the algebra $\Lambda(V,n)$ can be
equivalently defined by generators and relations:

\begin{equation*}
\left\{e_{i_1},\cdots,e_{i_n}\right\}=\sum \limits_{\tau \in S_n}
e_{i_{\tau(1)}}\cdots e_{i_{\tau(n)}}=0,
\end{equation*}

\noindent with $S_n$ the group of permutations with $n$ elements.
\end{remark}

If we define $\Lambda(V,n)_i$ the set of elements of degree $i \text { (mod. }
n$) we have $\Lambda(V,n)_i \Lambda(V,n)_j \subseteq \Lambda(V,n)_{i+j} \text{
(mod. } n)$ and

\begin{eqnarray}
\Lambda(V,n) = \Lambda(V,n)_0 \oplus \cdots \oplus \Lambda(V,n)_{n-1}
\end{eqnarray}

\noindent thus $\Lambda(V,n)$ is a $\mathbb{Z}_n-$graded vector space. When $%
n=2$, $\Lambda(V,2)$ coincides with the usual exterior algebra. But when $n
>2 $, $\Lambda(V,n)$ is very different from the exterior algebra. Indeed, $%
\Lambda(V,n)$ is defined through $n$-th order relations, and consequently
the number of independent monomials increases with polynomial's degree (for
instance, $(e_1 e_2)^k, k \ge 0$ are all  independent). This means that we
do not have enough constraints among the generators to order them in some
fixed way and, as a consequence, $\Lambda(V,n)$ turns out to be an 
infinite-dimensional algebra. In particular if we define $\Lambda_k(V, n)=
V^{\otimes^k}/\mathcal{I}(V,n)$, we have $\Lambda_0(V, n)=\mathbb{C},
\Lambda_1(V, n)= V, \cdots, \Lambda_{n-1}(V, n)=V^{\otimes^{n-1}}$, but the
spaces $\Lambda_k(V, n), k \ge n$ are not empty and more difficult to
characterise. In order to construct a basis of $\Lambda(V,n)$, \textit{i.e.}
to specify $\Lambda_k(V, n), k \ge n$ we need one more definition.

\begin{definition}
\label{rise} We say that the sequence $(i_1, i_2,\cdots i_k), \ k \ge n$
with $i_1,\cdots, i_k \in \left\{1,\cdots,d\right\}$ has a rise of length $n$
if it exists $0 \le \ell \le k-n$ such that $i_{\ell+1} \le i_{\ell+2}
\le \cdots \le i_{\ell+n}$. We denote by $I_{k,n} \subseteq
\left\{1,\cdots,d\right\}^k $ the set of $k$ indices which has a rise of
length $n$.
\end{definition}

\begin{theorem}
\label{roby} (N. Roby \cite{roby}) A basis of $\Lambda(V,n)$ is given by the
words $e_{i_1} \cdots e_{i_k}, k \in \mathbb{N}$ such that the
sequence $(i_1,\cdots, i_k)$ has not a rise of length $n$ \textit{i.e.} $%
(i_1,\cdots,i_k) \in \left\{1,\cdots,d\right\}^k\setminus I_{k,n}$. The
sequence $I=(i_1,\cdots,i_n)$ will be called a Roby sequence, and the
element $e_{_{\hskip -.05 truecm I}}=e_{i_1} \cdots e_{i_n}$ a Roby element
or a Roby word.
\end{theorem}

\begin{example}
\begin{enumerate}
\item If $\text{dim } V=2$ a basis of $\Lambda(V,3)$ is given by the
elements $t (e_2 e_1)^k t^{\prime }, k \in \mathbb{N}, t,t^{\prime
}=1,e_1,e_2$~\cite{roby}. This means that

\begin{eqnarray}
\Lambda_0(V,3)&=& \mathbb{C}  \notag \\
\Lambda_1(V,3)&=& V  \notag \\
\Lambda_2(V,3)&=& V \otimes V  \notag \\
\Lambda_{2n+1}(V,3) &=& \Big< e_i(e_2 e_1)^n, (e_2 e_1)^ne _i,\ i=1,2, \Big>%
, n \ge 1  \notag \\
\Lambda_{2n}(V,3) &=& \Big< e_i(e_2 e_1)^{n-1}e_j, \ i,j=1,2, (e_2 e_1)^n %
\Big>, n \ge 2.  \notag
\end{eqnarray}

\item If $\text{dim } V=4$ then $\text {dim } \Lambda_3(V,3) =44$, the
basis is given by

\begin{equation*}
\begin{array}{ll}
e_{i}e_{j}e_{i},\ e_{j}e_{i}e_{i}, & i<j, \\ 
e_{i}e_{i}e_{j},\ e_{i}e_{j}e_{i}, & i>j, \\ 
e_{j}e_{k}e_{i},\ e_{k}e_{i}e_{j},\ e_{i}e_{k}e_{j},\ e_{j}e_{i}e_{k},\
e_{k}e_{j}e_{i}, & i<j<k.%
\end{array}%
\end{equation*}

Similarly   $\text {dim } \Lambda_4(V,3) =256$.

\end{enumerate}
\end{example}

\medskip

\subsubsection{ A Poincar\'{e}-Birkhoff-Witt basis of $\mathcal{U}({%
\mathfrak{g}})$}\label{U-PBW}

Let $\mathfrak{g=g}_{0}\oplus \mathfrak{g}_{1}$ be an elementary Lie algebra
of order three.\ We denote by the letters $X$ (resp. $Y$) the elements of $%
\mathfrak{g}_{0}$ (resp. $\mathfrak{g}_{1}$). Then an element of $\mathcal{U}%
({\mathfrak{g}})$ is a word written using letters as $X_{\alpha }$ and $%
Y_{\beta}$ belonging to the free algebra $\mathcal{L}(X,Y)$ generated by $%
\{X_{1},\ldots ,X_{p},Y_{1},\ldots ,Y_{q}\},$ where $p=\dim \mathfrak{g}_{0}$
and $q=\dim \mathfrak{g}_{1}.$ We have the following three rules of
reduction of a word:

\begin{itemize}
\item $YX=XY-[X,Y].$ As $[X,Y]\in \mathfrak{g}_{1},$ this rule permits to
reduce a word as $m(X)m(Y)$ where $m(X)$ (resp. $m(Y)$) \ contains only $X$
(resp. $Y$).

\item $X_{2}X_{1}=X_{1}X_{2}-[X_{1},X_{2}].$ As $[X_{1},X_{2}]\in \mathfrak{g%
}_{0},$ this rule permits to write a word $m(X)$ on the classical Poincar%
\'{e}-Birkhoff-Witt words of $\mathcal{U}({\mathfrak{g}}_{0}).$

\item $%
Y_{a}Y_{b}Y_{c}=-Y_{b}Y_{a}Y_{c}-Y_{c}Y_{b}Y_{a}-Y_{a}Y_{c}Y_{b}-Y_{b}Y_{c}Y_{a}-Y_{c}Y_{a}Y_{b}-\{Y_{a},Y_{b},Y_{c}\}. 
$ If $\{Y_{a},Y_{b},Y_{c}\}=0$ for all $a,b,c,$ this rule permits to write a
word $m(Y)$ as a Roby word of the three-exterior algebra $\Lambda(%
\mathfrak{g}_{1},3).$
\end{itemize}

\bigskip The first two rules permits to write a word $m(X,Y)$ of the free
algebra $\mathcal{L}(X,Y)$ on the form $m_{1}(X)m_{2}(Y)$ with $m_{1}(X)\in 
\mathcal{U}(\mathfrak{g}_{0})$ $\ $and $m_{2}(Y)\in \mathcal{L}(Y)$. Then we
have to reduce $m_{2}(Y).$ Suppose that $m(Y)$ is a word of length $n$.\ We
will denote it by $m^{n}(Y).$ If this word is a Roby word, the reduction is
finished. If not, there exist in \ $m^{n}(Y)=Y_{a_{1}}Y_{a_{2}}\ldots
Y_{a_{n}}$ a triple $(a_{i} \le a_{i+1} \le a_{i+2})$ such that $%
(a_1,\cdots,a_n)$ is not a Roby sequence. In this case, the rule 3 permits
to write%
\begin{equation*}
m^{n}(Y)=\sum m_{i}^{n}(Y)+m^{n-2}(X,Y)
\end{equation*}%
where $m^{n-2}(X,Y)\in \mathcal{L}(X,Y)$ but it is of length $n-2.\ $Let $%
\mathcal{L}_{n-2}(X,Y)$ be the linear subspace of $\mathcal{L}(X,Y)$ whose
element are of length smaller or equal than $n-2.$ Then $m^{n}(Y)=\sum
m_{i}^{n}(Y)$ modulo $\mathcal{L}_{n-2}(X,Y).$ We can apply the Roby rule $3$
to obtain $m^{n}(Y)=\sum m_{\alpha }^{n}(Y)$ modulo $\mathcal{L}_{n-2}(X,Y)$
where $m_{\alpha }^{n}(Y)$ are Roby words. Now we have to consider words of
type $m^{n-2}(X,Y)$ of length $n-2$.\ The first two rules reduces such a
word to $m_{1}(X)m_{2}(Y)$ with $m_{1}(X)\in \mathcal{U}(\mathfrak{g}_{0})$ $%
\ $and $m_{2}(Y)\in \mathcal{L}(Y)$ but $m_{2}(Y)$ is a word of length
smaller than $n-2.$ An induction process permits to conclude.

\begin{theorem}
\label{PBW} The universal algebra is generated by a basis of the universal
enveloping algebra $\mathcal{U}(\mathfrak{g}_{0})$ and by a Roby basis of
the three-exterior algebra $\Lambda(\mathfrak{g}_{1},3)$. Then, as
vector space, we have%
\begin{equation*}
\mathcal{U}(\mathfrak{g})=\mathcal{U}(\mathfrak{g}_{0})\otimes \Lambda(%
\mathfrak{g}_{1},3).
\end{equation*}
\end{theorem}

\noi
This result has been conjectured in \cite{flie2}.\\

In the usual way, the representations of $\g$ 
are in bijective correspondence with the  representations
of the associative algebra ${\cal U}(\g)$.
Consequently, if  $ I \subset {\cal U}(\g)$ 
is a   two-sided ideal, then the quotient $ {\cal U}(\g)/{ I}$ gives 
a representation of $\g$. 
If $\g=\g_0 \oplus \g_1$, with $\g_0$  a semi-simple Lie algebra,
 we decompose $\g$ in its Borel
form
$\g= \n_+ \oplus \h \oplus \n_-=(\n_0{}_+ \oplus \h_0{}_+ \oplus \n_0{}_-)
\oplus (\n_1{}_+ \oplus \h_1{}_+ \oplus \n_1{}_-)$. 
If we assume that some highest weight representation can be obtained
 since the algebra
$\g$ is defined by cubic relations, 
 the representation so obtained will be infinite-dimensional.
This can be compare with the Roby Theorem \ref{roby} for the three-exterior
algebra.
This means that in finite-dimensional representations of $\g$,
as in the Example  \ref{mat}
 the elements of ${\mathfrak n}_-$
are not only nilpotent but also satisfy  additional relations and
the representation is non-faithful. See \cite{flie2} for more details. \\

\subsection{Example : ${\mathcal{U}}({\mathfrak{iso}}_3(1,3))$.}

We consider the Lie algebra of order three given in Example \ref{FP}.
Introduce $X_a=(L_{01}, L_{02}, L_{03}, L_{12}, L_{13}, L_{23},$ $P_0,P_1,
P_2, P_3)$ the generators of ${\mathfrak{iso}}(1,3)$ (we restrict ourselves
to $D=4$ since the generic case is analogous), and for any $\vec a=
(a_1,\cdots,a_{10}) \in \mathbb{N}^{10}$ set

\begin{eqnarray}  \label{b1}
X_{\vec a}= \frac{X_1^{a_1}}{a_1 !} \cdots \frac{X_{10}^{a_1}}{a_{10} !}.
\end{eqnarray}
Consider now $I_\ell=(i_{\mu_1}, \cdots, i_{\mu_\ell}) \in
\left\{0,1,2,3\right\}^\ell \setminus I_{\ell,3}$ a Roby sequence (see
Definition \ref{rise}) and define

\begin{eqnarray}  \label{b2}
V_{(\mu_1,\cdots,\mu_\ell)}= V_{\mu_1} \cdots V_{\mu_\ell}.
\end{eqnarray}
From Theorem \ref{PBW}, the Poincar\'e-Birkhoff-Witt basis is given by

\begin{eqnarray}  \label{b3}
\Big\{g_{\vec b, (\mu_1, \cdots, \mu_\ell)} = X_{\vec b}
V_{(\mu_1,\cdots,\mu_\ell)} \ , \ \vec b \in \mathbb{N}^{10}, \ell \in 
\mathbb{N}, (\mu_1,\cdots, \mu_\ell) \in \left\{0,1,2,3\right\}^\ell
\setminus I_{\ell,3}\Big\}.
\end{eqnarray}

\noindent Here we denote by $1$ the neutral element.\newline

The algebra structure constructed on $\mathcal{U}({\mathfrak{iso}}_{3}(1,3))$
is immediate, and for instance we have: $X_{\vec{b}}V_{(\mu_1,\cdots,\mu_%
\ell)}=g_{\vec{b},(\mu_1,\cdots,\mu_\ell)}$. The multiplication law for $X_{%
\vec{a}}$ and $X_{\vec{b}}$ uses explicitly the commutation relation of the
algebra ${\mathfrak{iso}}(1,3)$ and for instance $%
X_{a}X_{b}=X_{b}X_{a}+[X_{a},X_{b}]$ if $a>b$. But the multiplication rule
for $V_{(\mu_1,\cdots, \mu_\ell)}$ and $V_{(\nu_1,\cdots,\nu_{\ell^{\prime
}})}$ is more involved. \ Indeed, if $(\mu_1,\cdots,\mu_\ell)\in
\left\{0,1,2,3\right\}^{\ell } \setminus I_{\ell ,3}$ and $%
(i_{\nu_1},\cdots,\nu_{\ell^{\prime }})\in \left\{0,1,2,3\right\}^{\ell
^{\prime }}\setminus I_{\ell ^{\prime },3}$ this does not mean that $%
(i_{\mu_1},\cdots, i_{\mu_\ell}, i_{\nu_1},\cdots,i_{\nu_{\ell^{\prime
}}})\in \left\{0,1,2,3\right\}^{\ell+\ell^{\prime }} \setminus I_{\ell +\ell
^{\prime },3}$. For instance, since $(112)$ is not a Roby sequence, we have 
$V_{1}V_{(12)}=-\frac12 P_{2}-V_{(121)}-V_{(211)}$, for the algebra
of Example \ref{FP}. 

\medskip

\section{Hopf structure on $\mathcal{U(}\mathfrak{g)}$}

Let $\mathcal{U(}\mathfrak{g)}$ be the universal enveloping algebra
associated to $\g$ a Lie algebra of order three. We consider on $%
\mathfrak{g}$ a $\mathbb{Z}_{3}-$grading coming from the decomposition of
the structure of Lie algebra of order three. As $\mathfrak{g}$ is an
elementary Lie algebra of order three, we can consider $\mathfrak{g}_{0}$
with grading $0$ and $\mathfrak{g}_{1}$ with grading $1$.\
 Let $q$ be a cubic primitive root of unity. We consider
on $\mathcal{U(}\mathfrak{g)\otimes }\mathcal{U(}\mathfrak{g)}$ a structure
of algebra given by%
\begin{equation*}
(a\otimes c)\bullet (b\otimes d)=q^{\mid b\mid \mid c\mid }ab\otimes cd
\end{equation*}%
where $\mid x\mid $ denotes the grading of an homogeneous element $x%
\mathfrak{.~}$ This associative structure on $\mathcal{U(}\mathfrak{%
g)\otimes }\mathcal{U(}\mathfrak{g)}$ will be denoted by $\mathcal{U(}%
\mathfrak{g)}\underline{\mathfrak{\otimes }}\mathcal{U(}\mathfrak{g)}$. 
We consider
$$\overline{\Delta}:T(\mathfrak{g})\rightarrow \mathcal{U(}\mathfrak{g)}%
\underline{\mathfrak{\otimes }}\mathcal{U(}\mathfrak{g})
$$
defined by 
\begin{itemize}
\item $\overline\Delta (1)=1\otimes 1,$

\item $\overline\Delta (X)=1\otimes X+X\otimes 1,$

\item $\overline\Delta (Y)=1\otimes Y+Y\otimes 1,$
\end{itemize}

and

\begin{itemize}
\item $\overline\Delta (a\otimes b)=\sum q^{\mid a_{(2)}\mid \mid b_{(1)}\mid
}a_{(1)}b_{(1)}\otimes a_{(2)}b_{(2)}=\overline\Delta (a)\bullet \overline\Delta (b),$
\end{itemize}

\noi
for all $X\in \mathfrak{g}_{0},Y\in \mathfrak{g}_{1}$ and $a,b\in 
{T(}\mathfrak{g).}$

\begin{lemma}
If we consider $L_3(\mathcal{U(}\mathfrak{g)}\underline{\mathfrak{\otimes }}\mathcal{U(}\mathfrak{g)})$ 
the Lie algebra of order $3$ associated to the associative algebra
 $\mathcal{U(}\mathfrak{g)}\underline{%
\mathfrak{\otimes }}\mathcal{U(}\mathfrak{g)}$, then 
\begin{equation*}
\overline \Delta ([X,G])=[\overline \Delta (X),\overline \Delta (G)]_{L_3(\mathcal{U(}\mathfrak{g)}\underline{%
\mathfrak{\otimes }}\mathcal{U(}\mathfrak{g)})}
\end{equation*}%
and%
\begin{equation*}
\overline \Delta \left\{ Y_{1},Y_{2},Y_{3}\right\} =\{\overline \Delta (Y_{1}),\overline \Delta
(Y_{2}), \overline \Delta (Y_{3})\}_{L_3((\mathcal{U(}\mathfrak{g)}\underline{\mathfrak{%
\otimes }}\mathcal{U(}\mathfrak{g)})}
\end{equation*}%
with $X\in \mathfrak{g}_{0},$ $G \in {\mathfrak{g}}$ and $Y_{i}\in \mathfrak{%
g}_{1}$ and where $[,]_{L_3(\mathcal{U(}\mathfrak{g)}\underline{\mathfrak{%
\otimes }}\mathcal{U(}\mathfrak{g)})}$ and $\{,,\}_{L_3(\mathcal{U(}\mathfrak{g)}%
\underline{\mathfrak{\otimes }}\mathcal{U(}\mathfrak{g)})}$ denotes the
bracket and the symmetric product of the Lie algebra of order three,
$L_3(\mathcal{U(}%
\mathfrak{g)}\underline{\mathfrak{\otimes }}\mathcal{U(}\mathfrak{g)).}$
\end{lemma}
{\bf Proof } As $\mathcal{U(}\mathfrak{g)}\underline{\mathfrak{\otimes }}%
\mathcal{U(}\mathfrak{g)}\mathcal{\ }$\ is a graded associative algebra, we
consider the associated structure of Lie algebra of order three. Let us look
the second identity. We have $\overline \Delta (Y_{i})=1\otimes Y_{i}+Y_{i}\otimes 1.$
On $L_3(\mathcal{U(}\mathfrak{g)}\underline{\mathfrak{\otimes }}\mathcal{U(}%
\mathfrak{g)})$ we consider the product

\begin{equation*}
\{b_{1},b_{2},b_{3}\}_{L_3(\mathcal{U(}\mathfrak{g)}\underline{\mathfrak{\otimes 
}}\mathcal{U(}\mathfrak{g)})}=\Sigma _{\sigma \in S_{3}}b_{\sigma (1)}\bullet
b_{\sigma (2)}\bullet b_{\sigma (3)}\ ,\ \ \forall b_{1},b_{2},b_{3}\in 
\mathcal{U(}\mathfrak{g)}\underline{\mathfrak{\otimes }}\mathcal{U(}%
\mathfrak{g)}.
\end{equation*}%
We deduce%
\begin{equation*}  \label{compat0}
\{\overline \Delta (Y_{1}),\overline \Delta (Y_{2}),\overline \Delta (Y_{2})\}_{L_3(\mathcal{U(}\mathfrak{g)}%
\underline{\mathfrak{\otimes }}\mathcal{U(}\mathfrak{g))}}=\{1\otimes
Y_{1}+Y_{1}\otimes 1,1\otimes Y_{2}+Y_{2}\otimes 1,1\otimes
Y_{3}+Y_{3}\otimes 1\}_{L_3(\mathcal{U(}\mathfrak{g)}\underline{\mathfrak{%
\otimes }}\mathcal{U(}\mathfrak{g)})}.
\end{equation*}%
We have to compute the following products $\{1\otimes Y_{1},1\otimes
Y_{2},1\otimes Y_{3}\}_{L_3(\mathcal{U(}\mathfrak{g)}\underline{\mathfrak{%
\otimes }}\mathcal{U(}\mathfrak{g)})},$ $\{Y_{1}\otimes 1,Y_{2}\otimes
1,Y_{3}\otimes 1\}_{L_3(\mathcal{U(}\mathfrak{g)}\underline{\mathfrak{\otimes }}%
\mathcal{U(}\mathfrak{g)})},$ and the crossed products of the form $%
\{1\otimes Y_{1},1\otimes Y_{2},Y_{3}\otimes 1\}_{L_3(\mathcal{U(}\mathfrak{g)}%
\underline{\mathfrak{\otimes }}\mathcal{U(}\mathfrak{g)})},$ $\{1\otimes
Y_{1},Y_{2}\otimes 1,Y_{3}\otimes 1\}_{L_3(\mathcal{U(}\mathfrak{g)}\underline{%
\mathfrak{\otimes }}\mathcal{U(}\mathfrak{g)})}$ \textit{etc}. 
By definition%
\begin{equation*}
\{1\otimes Y_{1},1\otimes Y_{2},1\otimes Y_{3}\}_{L_3(\mathcal{U(}\mathfrak{g)}%
\underline{\mathfrak{\otimes }}\mathcal{U(}\mathfrak{g)})}=\Sigma _{\sigma
\in S_{3}}1\otimes Y_{\sigma (1)}\bullet 1\otimes Y_{\sigma (2)}\bullet
1\otimes Y_{\sigma (3)}=\Sigma _{\sigma \in S_{3}}1\otimes Y_{\sigma
(1)}Y_{\sigma (2)}Y_{\sigma (3)}=1\otimes \{Y_{1},Y_{2},Y_{3}\}.
\end{equation*}%
Similarly%
\begin{equation*}
\{Y_{1}\otimes 1,Y_{2}\otimes 1,Y_{3}\otimes 1\}_{L_3(\mathcal{U(}\mathfrak{g)}%
\underline{\mathfrak{\otimes }}\mathcal{U(}\mathfrak{g)})}=\Sigma _{\sigma
\in S_{3}}Y_{\sigma (1)}\otimes 1\bullet Y_{\sigma (2)}\otimes 1\bullet
Y_{\sigma (3)}\otimes 1=\Sigma _{\sigma \in S_{3}}Y_{\sigma (1)}Y_{\sigma
(2)}Y_{\sigma (3)}\otimes 1=\{Y_{1},Y_{2},Y_{3}\}\otimes 1.
\end{equation*}%
For the crossed terms, we have%
\begin{equation*}
\begin{array}{ll}
\{1\otimes Y_{1},1\otimes Y_{2},Y_{3}\otimes 1\}_{L_3(\mathcal{U(}\mathfrak{g)}%
\underline{\mathfrak{\otimes }}\mathcal{U(}\mathfrak{g)})} & =1\otimes
Y_{1}\bullet 1\otimes Y_{2}\bullet Y_{3}\otimes 1+1\otimes Y_{2}\bullet
1\otimes Y_{1}\bullet Y_{3}\otimes 1+Y_{3}\otimes 1\bullet 1\otimes
Y_{2}\bullet 1\otimes Y_{1} \\ 
& +1\otimes Y_{1}\bullet Y_{3}\otimes 1\bullet 1\otimes Y_{2}+1\otimes
Y_{2}\bullet Y_{3}\otimes 1\bullet 1\otimes Y_{1}+Y_{3}\otimes 1\bullet
1\otimes Y_{1}\bullet 1\otimes Y_{2} \\ 
& =(1+q+q^{2})(Y_{3}\otimes Y_{1}Y_{2}+Y_{3}\otimes Y_{2}Y_{1}) \\ 
& =0.%
\end{array}%
\end{equation*}%
Then 
\begin{eqnarray}  \label{compat}
\{\overline \Delta(Y_{1}),\overline \Delta (Y_{2}),\overline \Delta (Y_{2})\}_{L_3(\mathcal{U(}\mathfrak{g)}%
\underline{\mathfrak{\otimes }}\mathcal{U(}\mathfrak{g)})} &=&\{1\otimes
Y_{1},1\otimes Y_{2},1\otimes Y_{3}\}_{\mathcal{U(}\mathfrak{g)}\underline{%
\mathfrak{\otimes }}\mathcal{U(}\mathfrak{g)}}+\{Y_{1}\otimes 1,Y_{2}\otimes
1,Y_{3}\otimes 1\}_{\mathcal{U(}\mathfrak{g)}\underline{\mathfrak{\otimes }}%
\mathcal{U(}\mathfrak{g)}}  \notag \\
&=&1\otimes \{Y_{1},Y_{2},Y_{3}\}+\{Y_{1},Y_{2},Y_{3}\}\otimes 1  \notag \\
&=&\overline \Delta \{Y_{1},Y_{2},Y_{3}\}.
\end{eqnarray}%
Similarly, we prove that%
\begin{equation*}
\lbrack \overline \Delta (X_{1}),\overline \Delta (X_{2})]_{L_3(\mathcal{U(}\mathfrak{g)}\underline{%
\mathfrak{\otimes }}\mathcal{U(}\mathfrak{g)})}=\overline \Delta \lbrack X_{1},X_{2}].
\end{equation*}

\noindent Q.E.D. 

From this lemma, we deduce that $Ker(\overline \Delta)$ contains the ideal 
defining $\mathcal{U(}\mathfrak{g)}$. Then
applying Theorem \ref{univ},  $\overline\Delta$
induces a comultiplication

\begin{equation*}
\Delta :\mathcal{U(}\mathfrak{g)\rightarrow }\mathcal{U(}\mathfrak{g)}%
\underline{\mathfrak{\otimes }}\mathcal{U(}\mathfrak{g)}.
\end{equation*}
 The elements satisfying $\Delta (u)=1\otimes u+u\otimes 1$ are
called primitive. The comultiplication is coassociative 
\begin{equation*}
(\Delta \otimes Id)\circ \Delta =(Id\otimes \Delta )\circ \Delta .
\end{equation*}%
In fact $(\Delta \otimes Id)\circ \Delta (ab)=(\Delta \otimes Id)\circ
(\Delta (a)\bullet \Delta (b))=((\Delta \otimes Id)\circ \Delta (a))\bullet
((\Delta \otimes Id)\circ \Delta (b))$. This last relation comes from the
fact that the coproduct $\Delta $ is of degree $0$. It satisfies also 
$\Delta ([X,G])=\Delta (XG)-\Delta (GX)=1\otimes XG+X\otimes G+G\otimes
X+XG\otimes 1-1\otimes GX-G\otimes X-X\otimes G-GX\otimes 1$ 
that is 
\begin{equation*}
\Delta ([X,G])=[X,G]\otimes 1+1\otimes \lbrack X,G],
\end{equation*}%
for all $(X,G)\in \mathfrak{g}_{0}\times {\mathfrak{g}}$ that is $%
[X_{1},X_{2}]$ and $[X,Y]$ are primitive elements.\ Likewise, if $%
Y_{1},Y_{2},Y_{3}\in \mathfrak{g}_{1}$, then%
\begin{equation*}
\Delta \left\{ Y_{1},Y_{2},Y_{3}\right\} =\sum\limits_{\sigma \in
S_{3}}\Delta (Y_{\sigma (1)}Y_{\sigma (2)}Y_{\sigma (3)})=\left\{
Y_{1},Y_{2},Y_{3}\right\} \otimes 1+1\otimes \left\{
Y_{1},Y_{2},Y_{3}\right\}.
\end{equation*}%
Thus $\left\{ Y_{1},Y_{2},Y_{3}\right\} $ is also a primitive element.
This means that the elements of $\g$ are primitive elements.

The counity 
\begin{equation*}
\varepsilon :\mathcal{U(}\mathfrak{g)\rightarrow }\mathbb{C}
\end{equation*}%
is the morphism given from $\varepsilon (1)=1,$ $\varepsilon (X)=\varepsilon
(Y)=0$ and the antipode is the $\mathbb{Z}_3$ anti-homomorphism%
\begin{equation*}
S:\mathcal{U(}\mathfrak{g)\rightarrow }\mathcal{U(}\mathfrak{g)}
\end{equation*}%
given from 
\begin{equation*}
S(1)=1,\ S(X)=-X,\ S(Y)=-Y
\end{equation*}%
and satisfying%
\begin{equation*}
S(ab)=q^{|a||b|}S(b)S(a).
\end{equation*}

\begin{definition}
We call Hopf structure associated to a Lie algebra $\mathfrak{g}$ of order
three, the Hopf structure on the enveloping algebra $\mathcal{U(}\mathfrak{g)%
}$ given by $(\Delta ,\varepsilon ,S).$
\end{definition}

\bigskip

As a final remark for this subsection it should be mentioned that these
twisted Hopf algebras (defined by a twisted tensorial product) has been
introduced by Majid \cite{majid}  in order to describe $q-$deformations and
braiding structures. He called them anyonic Hopf algebras. It as to be
stressed that the structures we are considering have \textit{a priori}
nothing to do with braiding and $q-$deformations, even if they share some
similarities. It should also be mentioned that a different Hopf algebra
associated to Lie algebras of order three has been defined in \cite{ayu},
where the coproduct was defined by the usual tensorial product and the
twist (necessary to ensure \eqref{compat}) was generated by an additional
element of order three, the grading map $\varepsilon$. In fact these two
structures are related by the transmutation theorem that maps an anyonic
Hopf algebra to a an (untwisted) Hopf algebra \cite{majid}.

\bigskip

\begin{example}
For ${\mathcal{U}}({\mathfrak{iso}}_{3}(1,4))$ the coproduct is given by

\begin{eqnarray}  \label{cop1}
\Delta X_{\vec a} &=& \sum \limits_{\vec b + \vec c = \vec a} V_{\vec b}
\otimes V_{\vec c}  \notag \\
\Delta V_{I_\ell} &=& \sum \limits_{{I_{\ell'} +I_{\ell''}=I_\ell }}
 q^{-N((I{_{\ell^{\prime }}}%
,I_{\ell^{\prime \prime }}),I_\ell)} V_{I_{\ell^{\prime }}} \otimes
V_{I_{\ell^{\prime \prime }}},
\end{eqnarray}

\noindent where $I_\ell=(\mu_1,\cdots,\mu_\ell)\in
\left\{0,1,2,3\right\}^\ell \setminus I_{\ell, 3}$ and the sum is taken over
all complementary subsequences of $I_\ell$ $I_{\ell^{\prime}}=(\nu_1,\cdots,%
\nu_{\ell^{\prime }})$ and $I_{\ell^{\prime
\prime}}=(\rho_1,\cdots,\rho_{\ell^{\prime \prime }})$ with $%
\ell=\ell^{\prime }+\ell^{\prime \prime }$. In this sum, $N((I{%
_{\ell^{\prime }}},I_{\ell^{\prime \prime }}),I_\ell)$ represents the number
of successive transpositions which bring $(\nu_1,\cdots,\nu_{\ell^{\prime
}},\rho_1,\cdots,\rho_{\ell^{\prime \prime }})$ into the ordered sequence $%
(\mu_1,\cdots,\mu_\ell)$ transposing firstly $\rho_1$, then $\rho_2$ and
finally $\rho_{\ell^{\prime \prime }}$. For instance this factor becomes $1$
for $(111)(222) \to (121221)$ since the first $2$ jumps over two variables,
then the second $2$ jumps over one variable and the last $2$ does not jump.
However some care has to be taken since the R.H.S. of the second equation of %
\eqref{cop1} may contain some words which are not of the Roby type. Indeed,
it may happen that one of the two subsequences are not of the Roby type.
This means in particular that we have to use the rules of reduction of words
in order to obtain only elements in the Poincar\'e-Birkhoff-Witt basis. For
instance, if one calculate using \eqref{cop1} $\Delta(V_{1212})$ one obtains
terms like $V_{122}$ which are not of the Roby type, although for $%
\Delta(V_{11})$ and $\Delta(\Delta_{221})$ we only obtain Roby words. Using
the rule of reduction, in the first case, one finally obtains

\begin{eqnarray}  \label{no-roby}
\Delta V_{(11)}& =& V_{(11)} \otimes 1 + (1+q)V_1 \otimes V_1 + 1 \otimes
V_{(11)}  \notag \\
\Delta V_{(12)}&=&V_{(12)}\otimes 1 + V_1 \otimes V_2 + q V_2 \otimes V_1 +
1 \otimes V_{(12)} \\
\Delta V_{(221)}&=&V_{(221)} \otimes 1 + V_{22}\otimes V_1 + (q+q^2)
V_{(21)} \otimes V_2 + (q+q^2) V_{1} \otimes V_{(22)} + V_{2} \otimes
V_{(21)} +1 \otimes V_{(2^21)}.  \notag \\
\Delta(V_{1212})&=&V_{1212} \otimes 1 + (V_1 \otimes V_{212}-q V_2\otimes
V_{121}-q V_2 \otimes V_{211} - q^2 V_1 \otimes V_{212}-q^2 V_1 \otimes
V_{221} + V_2 \otimes V_{121})  \notag \\
&+&(V_{12}\otimes V_{12}+q V_{11} \otimes V_{22}+q^2 V_{12}\otimes V_{21}
+q^2 V_{21}\otimes V_{12}+V_{22}\otimes V_{11}+q V_{12} \otimes V_{12}) 
\notag \\
&+&(V_{121}\otimes V_2 -q V_{211} \otimes V_1 -q V_{221}\otimes V_1 -q^2
V_{121}\otimes V_2 -q^2 V_{211}\otimes V_2 + V_{212}\otimes V_1) + 1 \otimes
V_{1212}  \notag \\
&-&\frac12(q V_2\otimes P_2+q^2 V_1 \otimes P_1 + q P_1 \otimes V_1 + q^2
P_2 \otimes V_2)  \notag
\end{eqnarray}
\end{example}

\section{Dual of the Hopf algebra associated to Lie algebras of order three}
Having endowed the universal enveloping algebra with a Hopf algebra
structure, in this section we construct its dual. We mainly stress on the
algebra structure of ${\cal U}(\g)^*$ and define a natural coproduct.
\bigskip

\subsection{Associative algebra structure on $\mathcal{U}({\mathfrak{g}})^*$}

\bigskip Let ${\mathfrak{g}}$ be an elementary  Lie algebra of order three.
We have provided $\mathcal{U}(\mathfrak{g})$ with a Hopf algebra structure,
we would like to define now the Hopf dual $\mathcal{U}({\mathfrak{g}})^*$.
In the infinite-dimensional case the notion of dual of a Hopf algebra $H$ is
more involved. There is in fact two ways to define this notion. In the first
approach we restrict to some subset $H^\circ \subset H^*$ with the
appropriate properties. In the second approach we just focus on the pairing
(see \textit{e.g.} \cite{cp,majid}). Since we were able to construct a PBW
basis of $\mathcal{U}({\mathfrak{g}})$, we follow the second approach and 
we identify the generators of $\mathcal{U}({\mathfrak{g}})^*$. Let us
consider a PBW basis of $\mathcal{U}(\mathfrak{g})$ associated to a basis $%
\{X_{i},Y_{j}\}$ of ${\mathfrak{g}}$. We have seen that such a basis is
written as words $X_{I}Y_{J}$ where $X_{I}$ is a PBW word of $\mathcal{U}(%
\mathfrak{g})_{0}$ and $Y_{J}$ a Roby word. If $X_{I}Y_{J}$ is an element of
the basis of $\mathcal{U}(\mathfrak{g}),$ we denote by $\Psi ^{IJ}$ the
corresponding dual element \textit{i.e.} with the pairing $\Psi^{IJ}(X_K
Y_L)= \delta^I{}_K \delta^J{}_L$.  To simplify the notations we put $\alpha
^{I}=\Psi ^{I\varnothing }$ (\textit{i.e.} the dual vector of the word $%
X_{I})$, $\theta ^{J}=\Psi ^{\varnothing J}$ (the dual vector of the Roby
word $Y_{J}) $.

Let $\mathcal{U}(\mathfrak{g})^{\ast }$ be the dual vector space. The
following product 
\begin{equation*}
M:\mathcal{U}(\mathfrak{g})^*\times \mathcal{U}(\mathfrak{g}%
)^*\longrightarrow \mathcal{U}(\mathfrak{g})^*
\end{equation*}%
given by 
\begin{equation*}
M(f,g)(V)=\rho (f,g)\Delta (V)
\end{equation*}%
with 
\begin{equation*}
\rho (f,g)(u\otimes v)=f(u)g(v)
\end{equation*}%
defines an unitary associative algebra structure on $\mathcal{U}(\mathfrak{g}%
)^{\ast }$.  In particular we have 
\begin{equation*}
1(a)=\epsilon (a)
\end{equation*}%
for every $a$ in $\mathcal{U}(\mathfrak{g})$, where $\epsilon $ is the
counit of $\mathcal{U}(\mathfrak{g})$.

Moreover, if in the basis $\left< Z_{IJ}= X_I Y_J\right>$ the coproduct
writes

\begin{equation*}
\Delta Z_{IJ} = \delta_{IJ}{}^{KLMN} Z_{KL} \otimes Z_{MN}, 
\end{equation*}

\noindent in the dual basis we have

\begin{equation*}
M(\Psi^{IJ}, \Psi^{KL})= \delta_{MN}{}^{IJKL} \Psi^{MN}. 
\end{equation*}

\begin{proposition}
\label{dual-prod}
As a vector space we have the following isomorphism,

\begin{equation*}
\mathcal{U}({\mathfrak{g}})^* \cong \mathbb{C}[{\mathfrak{g}}_0] \otimes
\Lambda({\mathfrak{g}}_1,3), 
\end{equation*}

\noindent where  $\mathbb{C}[{\mathfrak{g}}_0]$ is the algebra of 
polynomials in $\mbox{dim}\ {\mathfrak{g}}_0$ variables and  $\Lambda({\mathfrak{g}}_1,3)$ the
three-exterior algebra in $\mbox{dim} \ {\mathfrak{g}}_1$ variables.
Moreover $\{\alpha^{i},\theta^{j}\},i=1,\ldots ,\dim \mathfrak{g}%
_{0},j=1,\ldots ,\dim \mathfrak{g}_{1}$ are the generators of the associative
algebra $\mathcal{U}(\mathfrak{g})^{\ast }$
\end{proposition}

{\bf Proof} We first prove that the zero-graded part is isomorphic to the
set of polynomials. Recall that $X_{\ell \ell} = \frac{X_\ell^2}{2}$ and for 
$\ell < s$ we have $X_{\ell s} = X_\ell X_s$. Since 
\begin{equation*}
\Delta (X_{\ell}X_{s})=X_{\ell}\otimes X_{s}+X_{s}\otimes X_{\ell}+
X_{\ell}X_{s}\otimes 1+1\otimes X_{\ell}X_{s},
\end{equation*}

\noindent we have%
\begin{equation*}
M(\alpha^{i_1},\alpha^{i_{2}})= \left\{ 
\begin{array}{ll}
\alpha^{i_{1}i_{2}} & i_1 < i_2, \\ 
\alpha^{i_2 i_1} & i_2 < i_1, \\ 
\alpha^{i_1 i_2} & i_1=i_2.%
\end{array}%
\right.
\end{equation*}

\noindent Furthermore, by induction we prove that%
\begin{equation*}
M(\alpha^{i_{1}},\alpha^{i_{2}},\ldots
,\alpha^{i_{p}})=\alpha^{i_{1}i_{2}\ldots i_{p}}
\end{equation*}%
as soon as $i_{1}<i_{2}<\ldots <i_{p}.$ Then the dual vectors $\alpha^{i},
i=1,\ldots ,\dim \mathfrak{g}_{0}$ generate the family $\left\{ \alpha^{I}
\right\}.$ It is also easy to prove by induction that the product in $%
\mathcal{U}({\mathfrak{g}}_0)^*$ is commutative. Indeed this last property
is simply a consequence of the cocommutativity of the coproduct in $\mathcal{%
U}({\mathfrak{g}}_0)$.\newline

We consider now the product in the graded sector. From 
\begin{equation*}
\Delta (Y_{\ell}Y_{s})=Y_{\ell}\otimes Y_{s}+qY_{s}\otimes Y_{\ell}+
Y_{\ell}Y_{s}\otimes 1+1\otimes Y_{\ell}Y_{s},
\end{equation*}%
we get%
\begin{equation*}
M(\theta^{j_{1}},\theta^{j_{2}})=\theta^{j_{1}j_{2}}+q\theta^{j_{2}j_{1}}
\end{equation*}%
for all $j_{1},j_{2}\in \{1,\ldots ,\dim \mathfrak{g}_{1}\}.$ We deduce that
the dual vectors $\theta^{j}$ generate the dual vectors of type $%
\theta^{j_{1}j_{2}}.$ Now we consider a Roby word $Y_{\ell m n }$ of length $%
3$. This implies that $(\ell m n)$ is a Roby sequence (we have not $\ell
\leq m \leq n).$ In this case

\begin{eqnarray*}
\Delta (Y_{\ell m n }) &=& Y_{\ell }Y_{m }Y_{n }\otimes 1 + Y_{\ell }Y_{m
}\otimes Y_{n } + qY_{\ell }Y_{n }\otimes Y_{m }+ q^{2}Y_{m }Y_{n }\otimes
Y_{\ell } \\
&+&Y_{\ell }\otimes Y_{m }Y_{n } +qY_{m }\otimes Y_{\ell }Y_{n } + q^{2}Y_{n
}\otimes Y_{\ell }Y_{m } +1\otimes Y_{\ell }Y_{m }Y_{n }.
\end{eqnarray*}

\noindent This means in particular that

\begin{equation*}
M(\theta^j,\theta^{jj})=0.
\end{equation*}

\noindent For $j_i < j_2$ we have

\begin{equation*}
\begin{array}{lll}
M(\theta^{j_1},\theta^{j_1 j_2}) & = & q^2 \theta^{j_1 j_2 j_1} \\ 
M(\theta^{j_1},\theta^{j_2 j_1}) & = & \theta^{j_1 j_2 j_1} - \theta^{j_2
j_1 j_1} \\ 
M(\theta^{j_2},\theta^{j_1 j_1}) & = & \theta^{j_2 j_1 j_1} + q \theta^{j_1
j_2 j_1};%
\end{array}
\end{equation*}

and for $j_1 \ne j_2 \ne j_3$ we obtain

\begin{equation*}
\begin{array}{llll}
M(\theta^{j_1},\theta^{j_2 j_3}) & = & \theta^{j_1 j_2 j_3} + q \theta^{j_2
j_1 j_3} +q^2 \theta^{j_2 j_3 j_1}, & \mbox{if~} (j_1j_2j_3), (j_2 j_1 j_3) %
\mbox{~and~} (j_2 j_3 j_1) \mbox{~are Roby sequences}, \\ 
M(\theta^{j_1},\theta^{j_2 j_3}) & = & q \theta^{j_2 j_1 j_3} +q^2
\theta^{j_2 j_3 j_1}, & \mbox{if~} (j_1j_2j_3) \mbox{~is not a  Roby
sequence}, \\ 
M(\theta^{j_1},\theta^{j_2 j_3}) & = & \theta^{j_1 j_2 j_3} +q^2 \theta^{j_2
j_3 j_1}, & \mbox{if~} (j_2 j_1 j_3) \mbox{~is not a Roby sequence}, \\ 
M(\theta^{j_1},\theta^{j_2 j_3}) & = & \theta^{j_1 j_2 j_3} + q \theta^{j_2
j_1 j_3}, & \mbox{if~} (j_2 j_3 j_1) \mbox{~is not a Roby sequence}.%
\end{array}
\end{equation*}
These formul\ae\ can be unified as follows we have $M(\theta^{\ell},%
\theta^{mn})= \theta^{\ell m n} + q \theta^{m \ell m} + q^2 \theta^{mn \ell}$%
, but in this last expression we just keep the terms corresponding to Roby
sequences. For instance if $(\ell m n)$ is not a Roby sequence we do not
have the first term in the sum above.  Thus, the dual vectors $\theta^j$
generate the vectors $\theta^{\ell m n}$ with $(\ell m n)$ a Roby sequence.
We now show that the vectors $\theta^j$ generates the three-exterior algebra
that is $M(\theta^\ell,\theta^m,\theta^n) + \mbox{perm.}=0$. We obviously
have $M(\theta^j,\theta^j,\theta^j)=0$.
(The multiplication in ${\cal U}(\g)^*$ is associative since
the coproduct in ${\cal U}(\g)$ is coassociative.) For $j_1 < j_2$, we have

\begin{equation*}
\begin{array}{lll}
M(\theta^{j_1},\theta^{j_1},\theta^{j_2}) & = & -\theta^{j_1 j_2 j_1}- q
\theta^{j_2 j_1 j_1}, \\ 
M(\theta^{j_1},\theta^{j_2},\theta^{j_1}) & = & 2 \theta^{j_1 j_2 j_1}-
\theta^{j_2 j_1 j_1}, \\ 
M(\theta^{j_2},\theta^{j_1},\theta^{j_1}) & = & -\theta^{j_1 j_2 j_1}- q^2
\theta^{j_2 j_1 j_1}, \\ 
&  & 
\end{array}
\end{equation*}

\noindent thus $M(\theta^{j_1},\theta^{j_1},\theta^{j_2}) + \mbox{perm.}=0$.
Finally, assume now that $j_1 < j_2 < j_3$ (this means that $(j_1 j_3 j_2),
(j_2 j_3 j_1), (j_2 j_1 j_3), (j_3 j_1 j_2)$ and $(j_3 j_2 j_1)$ are Roby
sequences although $(j_1 j_2 j_3)$ is not a Roby sequence), we have

\begin{eqnarray}
\begin{array}{lllllllllll}
M(\theta^{j_1}, \theta^{j_2}, \theta^{j_3}) & = & q^2 \theta^{ {j_2}{j_3}{j_1%
} } & + & q^2 \theta^{ {j_3}{j_1}{j_2} } & + & q \theta^{ {j_1}{j_3}{j_2} }
& + & q \theta^{ {j_2}{j_1}{j_3} } & + & \theta^{ {j_3}{j_1}{j_2} } \\ 
M(\theta^{j_2}, \theta^{j_3}, \theta^{j_1}) & = & \theta^{ {j_2}{j_3}{j_1} }
& + & q^2 \theta^{ {j_3}{j_1}{j_2} } & + & \theta^{ {j_1}{j_3}{j_2} } & + & 
q \theta^{ {j_2}{j_1}{j_3} } & + & q \theta^{ {j_3}{j_2}{j_1} } \\ 
M(\theta^{j_3}, \theta^{j_1}, \theta^{j_2}) & = & q^2 \theta^{ {j_2}{j_3}{j_1%
} } & + & \theta^{ {j_3}{j_1}{j_2} } & + & q \theta^{ {j_1}{j_3}{j_2} } & +
& \theta^{ {j_2}{j_1}{j_3} } & + & q \theta^{ {j_3}{j_2}{j_1} } \\ 
M(\theta^{j_1}, \theta^{j_3}, \theta^{j_2}) & = & \theta^{ {j_2}{j_3}{j_1} }
& + & q \theta^{ {j_3}{j_1}{j_2} } & + & \theta^{ {j_1}{j_3}{j_2} } & + & 
q^2\theta^{ {j_2}{j_1}{j_3} } & + & q^2 \theta^{ {j_3}{j_2}{j_1} } \\ 
M(\theta^{j_2}, \theta^{j_1}, \theta^{j_3}) & = & q \theta^{ {j_2}{j_3}{j_1}
} & + & \theta^{ {j_3}{j_1}{j_2} } & + & q^2 \theta^{ {j_1}{j_3}{j_2} } & +
& \theta^{ {j_2}{j_1}{j_3} } & + & q^2 \theta^{ {j_3}{j_2}{j_1} } \\ 
M(\theta^{j_3}, \theta^{j_2}, \theta^{j_1}) & = & q \theta^{ {j_2}{j_3}{j_1}
} & + & q \theta^{ {j_3}{j_1}{j_2} } & + & q^2 \theta^{ {j_1}{j_3}{j_2} } & +
& q^2\theta^{ {j_2}{j_1}{j_3} } & + & \theta^{ {j_3}{j_2}{j_1} }%
\end{array}
\notag
\end{eqnarray}

\noindent and thus $M(\theta^{j_1},\theta^{j_1},\theta^{j_3}) + \mbox{perm.}%
=0$.

To end the proof in the graded sector, we observe that the coproduct for
Roby words of length greater than three is given by the second equation of %
\eqref{cop1} eventually corrected by terms similar to those
appearing in the last equation of \eqref{no-roby} (when
in the right hand side there is terms which are not of the Roby type). This means
that if one calculates $M(\theta^{j_1}, \theta^{j_2 \cdots j_n})$ two types
of terms will be obtained

\begin{enumerate}
\item $\theta^{j_1 j_2 \cdots j_n} + q \theta^{j_2 j_1 \cdots j_n} + \cdots
+ q^{n-1}\theta^{j_2 \cdots j_n j_1}$, where in the previous summation the
non-Roby words are excluded;

\item terms of the type above coming from the reduction in the PBW basis of
the non-Roby words which appear in the coproduct \eqref{cop1}.
\end{enumerate}

This means that dual vectors  $\{\theta^{j}\}_{j=1,\ldots,\dim \mathfrak{g}%
_{1}\text{ }}$ generate the family $\theta^{J},$ and in particular that $%
<~\theta^J, J \mbox{~ Roby sequence}> $ is isomorphic to the three-exterior
algebra.\newline

Moreover the non-Roby terms in the coprodruct $\Delta V_J$ will induce a
product which mixes $\alpha-$ and $\theta-$types of terms. This means that
we have $M(\alpha^{i},\theta^{j})=\Psi^{ij} $ plus possibly some terms
involving $\theta^{j_1 j_2 j_3 j_4}$. Then $\{\alpha^{i},\theta^{j}\},i=1,%
\ldots ,\dim \mathfrak{g}_{0},j=1,\ldots ,\dim \mathfrak{g}_{1}$ are the
generators of the associative algebra $\mathcal{U}(\mathfrak{g})^{\ast }$.
It is important to notice that  the variables $\alpha^i$ and $\theta^j$
do not commute.
 For instance  for the algebra $\mathfrak{iso}_3(1,3)$
given in Example \ref{FP}, looking to $\Delta V_{1212}$ in
 \eqref{no-roby} shows explicitly that
$\theta^1 \alpha^1 \ne \alpha^1 \theta^1$. 
Thus  as a vector space we have the following 
 isomorphism $\mathcal{U}({\mathfrak{g%
}})^*\cong \mathbb{C}[{\mathfrak{g}}_0] \otimes \Lambda({\mathfrak{g}}_1,3).$
\noindent Q.E.D. \newline

\noindent{\bf Remarks. }

1. In order to respect the grading structure of the algebra, since the elements
of ${\mathfrak{g}}_1$ are of grade-one, we assume here that the variables $%
\theta^j$ are of grade two.

2. In \cite{ayu}, the vectorial description of $\mathcal{U}({\mathfrak{g}})^*$ does not correspond to the result of proposition 3.
In fact, in this paper, the authors do not use Roby algebras to describe the dual of the enveloping algebra.

\subsection{Hopf algebra structure on ${\cal U}(\g)^*$}

As the algebra $\mathcal{U}({\mathfrak{g}})$
 is graded, we can put a coalgebra structure on its dual vectorial. To define this coproduct, we consider
the vectorial dual basis $\{\Psi^{MN}\}$ of the PBW basis $\{Z_{IJ}\}$ of $\mathcal{U}({\mathfrak{g}})$. If we put
$$Z_{IJ}.Z_{KL}=\mu_{IJKL}{}^{MN}Z_{MN}$$ then the linear map
\begin{equation*}
\Delta : \mathcal{U}({\mathfrak{g}})^* \longrightarrow \mathcal{U}({%
\mathfrak{g}})^* \otimes \mathcal{U}({\mathfrak{g}})^* 
\end{equation*}
given by 
\begin{equation*}
\Delta (\Psi ^{MN})=\mu _{_{IJKL}}{}^{MN}\Psi ^{IJ}\otimes \Psi ^{KL}.
\end{equation*}
satisfies in particular
\begin{equation*}
(\Delta f)(X\otimes Y)=f(XY)
\end{equation*}%
for every $f\in \mathcal{U}({\mathfrak{g}})^{\ast }$ and $X,Y\in \mathcal{U}(%
{\mathfrak{g}})$. We deduce%
\begin{equation*}
(\Delta \otimes Id)\Delta (\Psi ^{MN})=\mu _{_{IJKL}}{}^{MN}\Delta (\Psi
^{IJ})\otimes \Psi ^{KL}=\mu _{_{IJKL}}{}^{MN}\mu _{_{RSTU}}{}^{IJ}\Psi
^{RS}\otimes \Psi ^{TU}\otimes \Psi ^{KL}
\end{equation*}%
and%
\begin{equation*}
(Id\otimes \Delta )\Delta (\Psi ^{MN})=\mu _{_{IJKL}}{}^{MN}\Psi
^{IJ}\otimes \Delta (\Psi ^{KL})=\mu _{_{IJKL}}{}^{MN}\mu
_{_{RSTU}}{}^{KL}\Psi ^{IJ}\otimes \Psi ^{RS}\otimes \Psi ^{TU}
\end{equation*}%
and the coassociativity is given by%
\begin{equation*}
\mu _{_{IJKL}}{}^{MN}\mu _{_{RSTU}}{}^{IJ}=\mu _{_{IJKL}}{}^{MN}\mu
_{_{RSTU}}{}^{KL}.
\end{equation*}

\noi
This last identity is just a consequence of the associativity of the product
in ${\cal U}(\g)$. This associativity being itself a consequence of
the independence of the way we reduce the non-Roby word in ${\cal U}(\g_1)$.

\medskip
However, we have seen that the PBW basis of ${\mathfrak{g}}_1$ (noted $B_1$)
is strongly related to the Roby elements $Y_I$, or to the Roby sequences $I$%
. This leads to a difficulty. Its is obvious that the set of Roby sequences
is not a representation of ${\mathfrak{g}}_0$. This means that we may obtain
results where the ${\mathfrak{g}}_0-$equivariance is not manifest. For
instance, if we consider the words of length three of the type $Y_{112},
Y_{121}, Y_{211}$ (\textit{i.e.} when two indices are equal) only the last
two are of the Roby type (since $1\le 1 \le 2$). However, if one chooses the
cubic 
elements of $B_1$ to be  $Y_{j_1 j_2 j_1}, Y_{j_2 j_1 j_1}$ with $j_1 < j_2$
and $Y_{j_1 j_1 j_2}, Y_{j_1 j_2 j_1}$ when $j_1 > j_2$ we will have a
result where the ${\mathfrak{g}}_0-$equivariance will not be manifest.
Indeed, it might happen that there is a $G_0$ (the Lie group of ${\mathfrak{g%
}}_0$) transformation that maps the Roby words $Y_{121}, Y_{211}$ to $%
Y_{212}, Y_{122}$. But the last word is not a Roby word. To solve this
problem, we take a modified Roby basis for which the ${\mathfrak{g}}_0-$%
equivariance would be manifest. Namely, we take the words $Y_{j_1 j_1 j_2},
Y_{j_2 j_1 j_1}, j_1 \ne j_2$ to be the modified Roby elements of $B_1$. We
clearly see that when $j_1=2,j_2=1$ the word $Y_{122}$ is not a Roby word in
the sense originally defined, but is a Roby word in the modified sense. This
rule is extended to words of length greater than three.

\begin{example}
For ${\mathcal{U}}({\mathfrak{iso}}_{3}(1,4))^*$ the coproduct is given by 
\begin{eqnarray}  \label{co-prod*}
\Delta x^\nu &=& x^\nu \otimes 1 + 1 \otimes x^\nu - f_{\alpha \beta,
\gamma}{}^\nu \alpha^{\alpha \beta} \otimes x^\gamma  \notag \\
&&\frac12\frac{1}{1-q^2}\eta_{\mu \rho} \theta^\mu \otimes (\theta^\nu
\theta^\rho-q \theta^\rho \theta^\nu) +\frac12\frac{1}{1-q^2}\eta_{\mu
\rho}(\theta^\rho \theta^\nu-q \theta^\nu \theta^\rho) \otimes \theta^\mu 
\notag \\ 
\Delta \alpha^{\mu \nu} &=& \alpha^{\mu \nu} \otimes 1 + 1\otimes
\alpha^{\mu \nu} +\frac12 F_{\alpha \beta, \gamma \delta}{}^{^{\mu \nu}}
\alpha^{\alpha \beta} \wedge \alpha^{\gamma \delta} \\
\Delta \theta^\mu &=& \theta^\mu \otimes 1 + 1 \otimes \theta^\mu -f_{\alpha
\beta, \gamma}{}^\mu \alpha^{\alpha \beta} \otimes \theta^\gamma,  \notag
\end{eqnarray}

\noindent with $f_{\alpha \beta, \gamma}{}^\mu $ and $%
F_{\alpha \beta, \gamma \delta}{}^{^{\mu \nu}}$ the structure constant defined
in Example \ref{FP} and $x^\mu, \alpha^{\mu \nu}, \theta^\mu$ the dual
vectors of $P_\mu, L_{\mu \nu}$ and $V_\mu$ respectively.
\end{example}

Finally the antipode and the counit are easily constructed. Indeed, if for
a Hopf algebra $H$ we have $S(e_i)= s_i{}^j e_j$, the pairing 
$\mathcal{U}({\mathfrak{g}})^*$ gives  for its
dual $S^*(e^i)= s_j{}^i e^j$. The counit of $H^\star$ is the unit of $H$.
Thus we have:

\begin{eqnarray}
\begin{array}{llll}
S^*(1) = 1, & S^*(x^\mu)= -x^\mu, & S^*(\omega^{\mu \nu}) = -\omega^{\mu
\nu}, & S^*(\theta^\mu)=-\theta^\mu \\ 
\epsilon^*(1) = 1, & \epsilon^*(x^\mu)= 0, & \epsilon^*(\omega^{\mu \nu}) =
0, & \epsilon^*(\theta^\mu)=0.%
\end{array}%
\end{eqnarray}

\begin{remark}
 Up to now we were considered complex
elementary Lie algebras of order three. However, such
algebras admit real forms. A real elementary Lie algebra of order three is
given by a real Lie algebra ${\mathfrak{g}}_0$ and ${\mathfrak{g}}_1$ a real
representation of $\g_0$ which satisfy the axioms of complex elementary Lie
algebras of order three. Of course when considering a real form of a complex 
${\mathfrak{g}}$ some structures are lost such that the grading map $%
\varepsilon$, the coproduct \textit{etc.} 
At a first glance it seems that the dual algebra
is also lost  since its product involves explicitly complex numbers.
However, we can forget the multiplication laws given previously and just
keep the fact that this algebra is simply given by the set of polynomials
and the three-exterior algebra. This is possible since $\Lambda({\mathfrak{g}%
}_1,3)$ can be defined consistently as a real algebra.
\end{remark}

\section{Conclusion and perspective}
In this paper we have studied on  formal ground
a particular class of ternary algebras named
Lie algebras of order three. 
We have then shown that 
one is able to associate to Lie algebras of order three a
corresponding universal enveloping algebra. A Poincar\'e-Birkhoff-Witt theorem
has been proven in this context. 
It turns out that the PBW basis is strongly related
to the three-exterior algebra (or the Roby algebra).  It has then been
 shown that
this universal enveloping algebra can be endowed with a Hopf algebra 
structure.  

 At that point one may wonder whether or not we may define
groups associated to Lie algebras of order three along the lines
one associates Lie groups to Lie algebras.
Indeed, if $G$ is a simply connected
Lie group, there is a duality between ${\cal U}(\g)$ (the universal
enveloping algebra of $\g$, the Lie algebra of $G$) and ${\cal F}(G)$
(the vector space of complex valued functions on $G$). This means that 
${\cal F}(G)$ is isomorphic to a subspace of ${\cal U}^*(\g)$ (the dual
of ${\cal U}(\g)$)\cite{cp,majid}. In other words one may wonder if a similar
construction applies in the context of Lie algebras of order three.
Incidently,  partial results have been obtained in this direction: 
matrix groups involving
 matrices with elements
belonging to the three-exterior algebra were defined in \cite{valla}.
The question of the relationship of these matrix groups and possible 
groups associated to Lie algebras of order three is still open.

\bigskip
\noi
{\bf Acknowledgments}
J. Lukierski is gratefully acknowledged for discussions and suggestions.


\begin{thebibliography}{99}
\bibitem{cm}
S.~Coleman and J.~Mandula, {\it Phys. Rev.} {\bf 159} (1967) 1251.
%
\bibitem{hls}
R.~Haag, J.~T.~Lopuszanski and M.~F.~Sohnius, {\it Nucl. Phys.} {\bf B88} 
(1975) 257.
%
\bibitem{bg1}
I. Bars   and  M. Gunaydin, 
  {\it  J.\ Math.\ Phys.\ }  {\bf 20} (1979)  1977.
\bibitem{bg2}
I. Bars I  and M.  Gunaydin, 
{\it   Phys.\ Rev.\  } {\bf D22}  (1980)  1403.
%
\bibitem{k1}
L. Vainerman  and R. Kerner, {\it J. Math. Phys}
{\bf 37}  (1995)  2553.
%
\bibitem{k2}
R. Kerner, {\it Class. and Quantum Grav.} {\bf 14 (1A)} (1997) A203.
\bibitem{k3}
A.  Borowiec,  N. Bazunova   and 
R. Kerner  2004 {\it  Lett. Math. Phys.}
{\bf  67}   (2004) 195.
%
\bibitem{r}
M. Rausch de Traubenberg,
 {\it Clifford algebras, supersymmetry and 
$\mathbb Z_n-$symmetries: Applications in
field theory,}
  arXiv:hep-th/9802141 (Habilitation Thesis).
%
\bibitem{f}
V. T. Filipov,    {\it Sibirsk. Math. Zh} {\bf 26} (1985) 126.
%
\bibitem{g}
A. V. Gnedbaye,    {\it C. R. Acad. Sci.  Paris S\'er. I Math.}
{\bf   321}  (1995)  147.
%
\bibitem{mv}
P. W. Michor   and  A. M. Vinogradov,   
{\it Rend. Sem. Mat. Univ. Politec. Torino} {\bf   54} (1996) 373.
%
\bibitem{gr}
N. Goze and E. Remm, {\it On n-ary algebras given by Gerstenhaber's products},
arXiv:0803.0553[math.RA]. 
%
\bibitem{bl}
 J.~Bagger and N.~Lambert,
  {\it Phys.\ Rev.\ }  {\bf D75}, (2007) 045020,
  arXiv:hep-th/0611108.
%
\bibitem{flie1} 
M.~Rausch de Traubenberg and M.~J.~Slupinski, 
{\it J.\ Math.\ Phys.\ } \textbf{%
41} (2000) 4556, arXiv:hep-th/9904126. 
%
\bibitem{flie2} M.~Rausch de Traubenberg and M.~J.~Slupinski, 
{\it J.\ Math.\ Phys. }\ \textbf{43}
(2002), 5145 arXiv:hep-th/0205113. 
%
\bibitem{flie3} M.~Goze, M.~Rausch de Traubenberg and A.~Tanasa, 
{\it J. \ Math. \ Phys. } {\bf 48} (2007) 093507,
arXiv:math-ph/0603008. 
%
\bibitem{anyons}
M.~Rausch de Traubenberg and M.~Slupinski,
{\it Mod. Phys. Lett. } {\bf A12} (1997) 3051, hep-th/9609203.
%
\bibitem{cubic1} N.~Mohammedi, G.~Moultaka, M.~Rausch de Traubenberg, 
{\it Int.\ J.\ Mod.\ Phys.\ }  {\bf A19} (2004) 5585, arXiv:hep-th/0305172. 
%
\bibitem{cubic3} G.~Moultaka, M.~Rausch de Traubenberg, A.~Tanasa, 
{\it Int. J. Mod. Phys.} {\bf A20}
(2005) 5779, arXiv:hep-th/0411198.
%
\bibitem{pform}
G. Moultaka, M. Rausch de Traubenberg  and A. Tanasa,
 { Proceedings of the XIth International Conference Symmetry Methods 
in Physics, Prague 21-24 June 2004}, arXiv:hep-th/0407168.
%
\bibitem{noether} M.~Rausch de Traubenberg, 
 {\it Pr. Inst. Mat. Nats. Akad. Nauk Ukr. Mat. Zastos.}, 50, Part
1, 2, 3, Natsional. Akad. Nauk Ukra\"{\i}ni, \~{I}nst. Mat., Kiev, 2004, pp.
578-585, arXiv:hep-th/0312066. 
%
\bibitem{grt1}
M. Rausch de Traubenberg,  
{\it Phys. Atom. Nucl. } {\bf 71} (2008) 1102, 
 arXiv:hep-th/0612204.
%
\bibitem{cr}
R.~Campoamor-Stursberg and M.~Rausch de Traubenberg,
  {\it J.\ Math.\ Phys.\  } {\bf 49} (2008) 063506,
  arXiv:0801.2630 [hep-th].

%
\bibitem{cp} V. Chari and A. Pressley, \textit{A Guide to Quantum Groups},
Cambridge University Press, Cambridge, 1995.
%
\bibitem{majid} S. Majid, \textit{Foundations of Quantum Group Theory},
Cambridge University Press, Cambridge, 1995.
%
%
\bibitem{roby} N. Roby, 
 {\it Bull. Sc. Math. } \textbf{94} (1970) 49.
%
%
\bibitem{valla}
M.~Rausch de Traubenberg,
{\it J. Phys. Conf. Ser. } {\bf 128} (2008) 012060, 
  arXiv:0710.5368 [math-ph].
%
\bibitem{ayu} H.~Ahmedov, A.~Yildiz and Y. Ucan, 
{\it J.\ Phys.} {\bf A 34} (2001) 6413,
arXiv:math.RT/0012058. 
%
\bibitem{cliff} N.~Roby, 
 {\it C. R. Acad. Sc. Paris}, \textbf{268} (1969) A484;
%
 Ph.~Revoy, 
{\it C. R. Acad. Sc. Paris}  \textbf{284} (1977)
A985; 
%
N.~Fleury, M.~Rausch de Traubenberg, 
{\it J. Math. Phys. } \textbf{33} (1992) 3356;
%
 N.~Fleury and
M.~Rausch de Traubenberg, 
{\it Adv. Appl. Cliff. Alg. } \textbf{4} (1994)
123.
%
%
%
%
%
\end{thebibliography}
\end{document}